\documentclass[a4paper,oneside]{article}
\usepackage{graphicx}
\def\e{\varepsilon}

\newcommand{\be}{\begin{equation}}
\newcommand{\ee}{\end{equation}}
\newcommand{\ba}{\begin{array}}
\newcommand{\ea}{\end{array}}

\def\Black{}

\def\circa#1{\,\raise.3ex\hbox{$#1$\kern-.75em\lower1ex\hbox{$\sim$}}\,}
\makeatletter
\def\art{\@ifnextchar[{\eart}{\oart}}
\def\eart[#1]#2#3#4#5#6{{\rm #2}, {\em #3 \bf #4} {\rm (#6) #5} ({\em #1})}
\def\hepart[#1]#2{{\rm #2, \em#1}}
\newcommand{\oart}[5]{{\rm #1}, {\em #2 \bf #3} {\rm (#5) #4}}

\newcounter{alphaequation}[equation]
\def\thealphaequation{\theequation\hbox to
0.6em{\hfil\alph{alphaequation}\hfil}}
\def\eqnsystem#1{
\def\@eqnnum{{\rm (\thealphaequation)}}
\def\@@eqncr{\let\@tempa\relax \ifcase\@eqcnt \def\@tempa{& & &} \or
  \def\@tempa{& &}\or \def\@tempa{&}\fi\@tempa
  \if@eqnsw\@eqnnum\refstepcounter{alphaequation}\fi
\global\@eqnswtrue\global\@eqcnt=0\cr}
\refstepcounter{equation} \let\@currentlabel\theequation \def\@tempb{#1}
\ifx\@tempb\empty\else\label{#1}\fi
\refstepcounter{alphaequation}
\let\@currentlabel\thealphaequation
\global\@eqnswtrue\global\@eqcnt=0 \tabskip\@centering\let\\=\@eqncr
$$\halign to \displaywidth\bgroup \@eqnsel\hskip\@centering
$\displaystyle\tabskip\z@{##}$&\global\@eqcnt\@ne
\hskip2\arraycolsep\hfil${##}$\hfil& \global\@eqcnt\tw@\hskip2\arraycolsep
$\displaystyle\tabskip\z@{##}$\hfil
\tabskip\@centering&\llap{##}\tabskip\z@\cr}
\def\endeqnsystem{\@@eqncr\egroup$$\global\@ignoretrue} \makeatother

\begin{document}
\centerline{hep-ph/0111373 \hfill INFN/TH-01/03}
\vspace{5mm}
\Black
\vspace{0.5cm}
\centerline{\LARGE\bf A Statistical Approach to}\vskip2mm
\centerline{\LARGE\bf Leptonic Mixings and Neutrino 
Masses\footnote{Presented 
in the poster session of the 11$^{th}$ Baksan School, 
April 18-24, 2001}}
\medskip\bigskip\Black
  \centerline{\large\bf  Francesco Vissani}\vspace{0.2cm}
  \centerline{\em INFN, Laboratori Nazionali del Gran Sasso,
Theory Group, I-67010 Assergi (AQ), Italy}

\vspace{1cm}
\centerline{\large\bf Abstract}
\begin{quote}\large\indent
Based on existing data, we argue for a peculiar structure 
of the neutrino mass matrix, that has a 
block of relatively large elements--a dominant block. 
We analyze this ansatz and extract its predictions, 
assuming that the ${\cal O}(1)$ coefficients (that are 
needed to define the model fully) are random variables. 
Further insights are obtained by postulating that 
this structure of the mass matrix is due to U(1) selection rules, 
a la Froggatt and Nielsen. A particularly interesting 
case emerges, when the angle $\theta_{13}$ (=the mixing $U_{\rm e3}$)
is within reach for next generation experiments, and the large mixing angle
solution for solar neutrinos is the preferred one.

\Black
\end{quote}
\vspace{5mm}

\section{Surprising Features of Massive Neutrinos}

\subsection*{Generalities}
Let us begin to recall certain general facts on neutrinos
\vskip-2mm
\begin{center}
\begin{tabular}{l|clc|r}
LEP &\ \ \  &$\nu_a$ & \ \ \  &
 3 active (=interacting) $\nu$'s \\
\hline
Big-bang N.S.\ \ \ \ \ & &$\nu_a,\ \nu_s$ & &
\ $\le 4$ $\nu$'s in thermal equilibrium
\end{tabular}
\end{center}

\noindent where we use the notation: 
$\nu=$ generically, a neutrino (or antineutrino);
$\nu_s=$ a sterile (=non-interacting) $\nu$-state;
$\nu_\mu=$ muon neutrino, {\em etc.};
$\nu_a=$ anyone among 
$\nu_{\rm e},\nu_\mu,\nu_\tau$ (active state 
are not distinguished by neutral current interactions--NC in the following).
Here  a list of observations that suggest oscillation:
\vskip-1mm
\begin{center}
\begin{tabular}{l|r|l|l}
atm-$\nu$ &
$
\left\{
 \begin{array}{l}
\nu_\mu \\
\nu_{\rm e} \\
\nu_a 
\end{array}
\right. 
$
&
$
\begin{array}{l}
- \\
= \\
+ 
\end{array}
$
& 
$
\begin{array}{l}
\mbox{especially low }E_\nu,\mbox{ large }L\\
\mbox{checked at reactors}\\
\mbox{Super-Kamiokande (SK) NC data}
\end{array}
$
\\ \hline 
sol-$\nu$ &
$
\left\{
\begin{array}{l}
\nu_{\rm e} \\
\nu_a 
\end{array}
\right. 
$
&
$
\begin{array}{l}
- \\
+
\end{array}$
& 
$
\begin{array}{l}
E_\nu\mbox{ dependence only in total rates} \\
\mbox{SK+SNO} 
\end{array}
$
\\ \hline 
LSND-$\nu$ &
$\stackrel{(-)}{\nu_{\rm e}}$ 
&  \, + 
& 
waiting for independent confirmation\\ \hline 
SN1987A-$\nu$ &
$\overline{\nu}_{\rm e}$
& =? & just 19 events; theoretical uncertainties
\end{tabular}
\end{center}

\noindent
{\em 1}$^{st}$ column, $\nu$-experiment (symbolical);
{\em 2$^{nd}$}, pertinent type of neutrino;
{\em 3$^{rd}$} column, 
what is presumably occuring, if
disappearance ``$-$''
of that type of neutrino, or appearance ``$+$'',
or neither of them ``$=$''; 
{\em e.g.} there is no claim for disappearance 
of atmospheric $\nu_{\rm e}.$ {\em 4$^{th}$} column, some comments. 
Note: Two cases for appearance are made by NC-, one 
by CC-events; all with similar significance, $\sim 3 \sigma.$
We proceed to comment on the two strongest ``anomalies''.

\subsection*{Atmospheric Neutrinos \& CHOOZ}

Super-Kamiokande has made a strong case for oscillations
with large mixing with:
$$\Delta m^2_{atm}=(1.5-5)\times 10^{-3} \mbox{ eV}^2$$
Their results are supported by MACRO and SOUDAN2.
In particular the quality of data is so high that 
in these experiments  $L/E_\nu$  modulation is visible,
and the hypothesis of oscillation of $\nu_\mu$
into a sterile state is strongly disfavored.
(Remaining doubts are connected to calculated $\nu$ fluxes, 
constraints of new cosmic ray data, hadronic uncertainties, 
and Baksan results).

Few  models have been concocted, aimed at reproducing 
some features of these data; but the
simplest explanation of a big set of data is 
almost pure  $\nu_\mu\to\nu_\tau$ oscillations:
$$
\theta_{23}=(45 \pm 10)^\circ\ \ \mbox{and} \ \ \ \theta_{13}<10^\circ
$$
The result on $\theta_{13}$ is  merit 
of the reactor experiment CHOOZ. When compared with quark mixing, 
such a big mixing is rather surprising.

\subsection*{Solar Neutrinos}

\begin{itemize}
\item The evidence for non-standard physics is
compelling (e.g.\ GALLEX/GNO and SAGE are 5 $\sigma$ away from
expected values). 
It is natural to assume that this is a manifestation of neutrino masses,
as for atmospheric $\nu,$ with $\Delta m^2_{sol}\ll \Delta m^2_{atm}.$

\item 
Total $\nu$-counting rates (with Standard Solar Model)
point to  certain ``solar $\nu$ solutions'',
with shorthands LMA, LOW, VO, QVO, SMA. 

\item 
Differential $\nu$-counting rates at
Super-Kamiokande give exclusion regions:
this  ``negative evidence'' is one reason why LMA
(the large mixing angle solution with 
$\theta_{12}\in [21^\circ,41^\circ]$) is 
favored in existing analyses. However, first SNO results
reinforce this inference. Unfortunately, the day-night
signal at SK is just a 1.5 $\sigma$ effect.\footnote{SN1987A 
electron anti-neutrino signals favor as little solar mixing as 
possible for LMA, together with certain values of $\Delta m^2_{sol}.$}

\end{itemize}

It is used to say that ``neutrinos are for patient people'',
but it seems that SNO NC data, together with 
Borexino/KamLAND results on longer term will satisfy even 
the impatient ones...

\subsection*{The Scandal of LMA}

This solution points to unexpected
$\nu$ 
properties, not only because 
$\theta_{12}$ is $2-3$ times  larger than the Cabibbo angle 
$\theta_C,$  but
also because of the weak ``hierarchy'':
$$\Delta m^2_{21}/\Delta m^2_{31}\sim 1/20-1/100$$
(compare it with charged fermion analogues).
This flurry of large mixings and weak hierarchies leads us to wonder: 
\vskip2mm
\centerline{{\fbox{\sf WHAT ARE NEUTRINOS TELLING US?}}}
\vskip2mm
\noindent In a few pages, we will see 
some guesswork on this point.

\section{Arguments for a Dominant Block}

\subsection*{Five Assumptions}
Here are the ingredients we use:
{\begin{enumerate}
\item There are {3 $\nu$} that mix among them. This explains
solar and atmospheric flux deficits. 
By def., {$m_1< m_2< m_3.$}
\item {LSND} has 3.2 $\sigma$ signal, but before interpretation
we wait for confirmation.
\item There is a bunch of {``small parameters''},
{
$$
\left\{
\begin{array}{l}
(\Delta m^2_{sol}/\Delta m^2_{atm})^{1/2} \\
|U_{\rm e3}|\sim\theta_{13}\\
|U_{\rm \mu 3}^2|-1/2\sim \theta_{23}-\pi/4
\end{array}\right.
$$}
let's term them collectively {$\varepsilon$} (adding a bit of
prejudice). 
\item The neutrino mass {spectrum} 
``resembles'' the usual ones,\footnote{This  hypothesis 
saves us from the need of operating a fine-tuning 
on a certain mixing.   In fact, SN1987A 
$\overline{\nu}_{\mu},$ 
$\overline{\nu}_{\tau}$ 
were probably not converted much  to $\overline{\nu}_{\rm e},$
since the measured energy is already quite low
when compared with expectations.} 
namely
$m_2-m_1\ll m_3-m_2$ to be contrasted
with  the possibility that 
$m_2-m_1\gg m_3-m_2$ (``inverted'' spectrum): 
$\Rightarrow$ 
$\Delta m^2_{21}=\Delta m^2_{sol}$
and 
$\Delta m^2_{31}=\Delta m^2_{atm}.$
\item The mass  {$m_1$} is not large\footnote{This
hypothesis saves us from the need of
fine-tunings: if we play to increase
$m_1,$ we have to tune more and more the mass differences,
since $m_j-m_i\sim \Delta m^2_{ji}/(2 m_1).$} in comparison with the 
smallest oscillation scale, {$\sqrt{\Delta m^2_{sol}}$}
\end{enumerate}}
Admittedly, this is quite a heavy mix of solid information and
prejudice--though, all assumptions seem, at least, defensible.

\subsection*{Inferring the Existence of a Dominant Block}
Let us begin by including only the biggest mass scale 
$m_3\sim(\Delta m^2_{atm})^{1/2}$ in ${\bf M_\nu}:$
$$
{\bf M_\nu}=m_3\ v_3\otimes v_3\ \ \ 
\mbox{with }v_3\approx (\e,1,1)/\sqrt{2}
$$
This, taken literally, implies:
\begin{equation}
{\bf M_\nu}\propto
\left(
\begin{array}{ccc}
\e^2 & \e & \e \\
\e & 1 & 1 \\
\e & 1 & 1
\end{array}
\right)
\label{db}
\end{equation}
\noindent Here is the ``dominant block''!  Adding 
{$m_2\ v_2\otimes v_2$} and {$m_1\ v_1\otimes v_1$}
modifies the elements of the matrix by terms order {$\e$}
and lifts the determinant of the ``dominant block'' from 0. 
Actually, it might be that the element {${\bf (M_\nu})_{\rm ee}$}
remains {${\cal O}(\e^2)$}, if the two little contributions 
tend to compensate each other, due to Majorana phases. However, it
is clear that at this level we are saying little on solar 
neutrinos, though we may naturally incorporate their oscillations.

\subsection*{Can we Weaken the Assumptions?}
How far can we go if we want to describe
{\em just} the atmospheric neutrino oscillations?
We  would like to argue that,
formally, we could say little on the neutrino 
mass matrix.\footnote{Though,  we could still get the dominant block 
renouncing to explain solar neutrinos, 
but maintaining the assumptions that the spectrum 
is not ``inverted'' there is no $m_1$ offset, 
and $m_3$ is the biggest mass scale.}
This is quite evident, after trying to imagine what
these mass matrices (in eV) have in common:
$$
10^{-2}\times \left(
\begin{array}{cc}
2.36 & 2.71 \\
2.71 & 3.12
\end{array}
\right), \ \ 
10^{-2}\times \left(
\begin{array}{cc}
0 & 7.27 \\
7.27 & 2.04
\end{array}
\right), 
$$
$$
\left(
\begin{array}{cc}
-9/65 & 91/92 \\
91/92 & 7/50
\end{array}
\right), \ \ 
1\!\! 1+ 10^{-3}\times
\left(
\begin{array}{cc}
20/31 & 20/27 \\
20/27 & 29/34
\end{array}
\right); 
$$
you may check that they all
have $\Delta m^2=3\times 
10^{-3}$ {\rm eV}$^2$
and $\theta=41^\circ.$ Thus, they could be not distinguished
by even an ideal atmospheric neutrino experiment, and 
no doubt that we are not in the ideal situation.\footnote{They 
differ because of parameters that are irrelevant 
to oscillations: $m_1$ and the Majorana phases.}
\vskip3mm
In other words, we are quite far from complete information!!! 
Or, from another point of view, there is space for speculations (theory).
For the reasons explained above, we will start from eq.\ (\ref{db}).

\section{The Meaning of Neutrino Mass Matrices with a Dominant Block}
\subsection*{An ``Electronic'' Selection Rule}
We assume that the structure of mass matrix (\ref{db})
is dictated by a selection rule, that requires  that
the elements with electron flavor have to pay some 
suppression factor $\e:$
\begin{equation}
{\bf M_\nu}\stackrel{{\cal O}(1)}{=}
\frac{\langle H\rangle^2}{M_X}
\left(
\begin{array}{ccc}
\e^2 & \e & \e \\
\e & 1 & 1 \\
\e & 1 & 1
\end{array}
\right) 
\label{db2}
\end{equation}
where  {$M_X=(0.8-1.6)\times 10^{14}$ GeV,} 
{$\langle H \rangle=174$ GeV},
and there is a bunch of {${\cal O}(1)$} coefficients.
There are some important qualifications:\\
\noindent $\bullet$ This is a {\em class} of mass matrices.\\
\noindent $\bullet$ The mass scale is fixed by hand; but 
{\em adimensional quantities} can be
predicted.\\
\noindent $\bullet$ The 
${\cal O}(1)$ complex coefficients can be specified assigning
a {random} phase, and a modulus=$1\, \pm$ 20 \% .\\ 
The last point is the most important. It means that we do not 
pretend to understand the details of the underlying theory; we 
concentrate on the ``gross'' structure.

\subsection*{The Underlying Mass Mechanisms}

Previous mass matrix might be due to the {\em vev} of a scalar 
{triplet} $\Delta,$ 
with family dependent couplings to leptons,
or \underline{even} to {seesaw} mechanism:
{
$$
{\bf M_\nu}=\langle H\rangle^2 {\bf Y_\nu} {\bf M}_R^{-1} {\bf Y}_\nu^t
$$
Proof: The hypothesis of family dependent couplings reads:\\
{${\bf Y_\nu}=\mbox{diag}(\e,1,1)\ {\cal O}(1)\
\mbox{diag}(\e^{n_1},\e^{n_2},\e^{n_3})$}
{and}
{${\bf M}_R=\mbox{diag}(\e^{n_1},\e^{n_2},\e^{n_3})\ 
{\cal O}(1)\ 
\mbox{diag}(\e^{n_1},\e^{n_2},\e^{n_3})$}\\
{
$\Rightarrow$
the powers of $\e$ attached to ``right-handed'' neutrinos
cancel in the light $\nu$ mass matrix 
(not in all observables).

\vskip3mm
\noindent Beware of ${\cal O}(1)$ matrices!} ${\cal O}(1)^{-1}\neq {\cal O}(1)$
$\Rightarrow$ ``triplet'' and ``seesaw'' yield 
\underline{different} outcomes.

\subsection*{A Check with Phenomenology}

Now, what remains to be done is, basically, 
to toss the dices and wish that the model is  successful.  
In next plot, we show the percentage of success 
as a function of $\e.$
\vskip7mm
\hbox{\hskip1.2cm
\includegraphics[width=9cm]{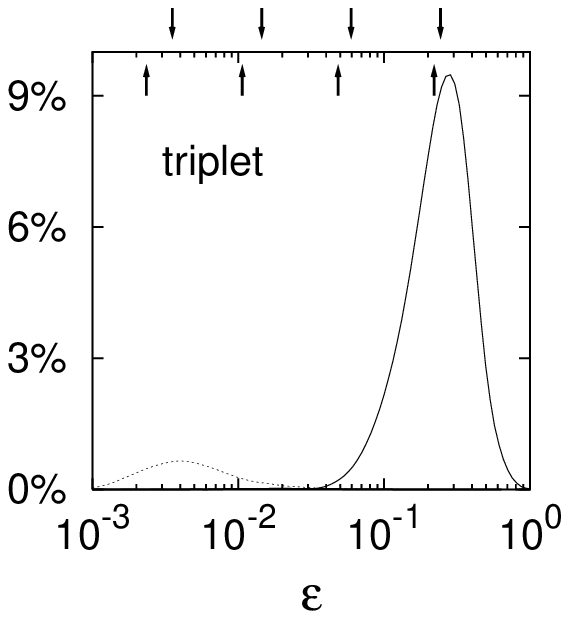}
\hskip-1.9cm
\includegraphics[width=9cm]{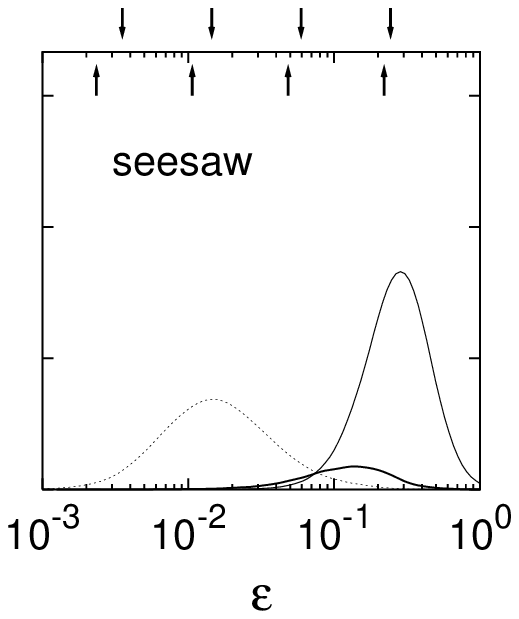}\hskip10cm.}
\noindent {\it Dashed line corresponds to SMA region,
continuous thin line to LMA, thick one to LOW.\\
We emphasize certain special values of $\varepsilon:$
$\e=(m_\mu/m_\tau)^{0.5,1,1.5,2}$, 
arrows pointing downward; $\e=(\sin\vartheta_C)^{1,2,3,4}$, 
arrows pointing upward.}
\vskip1mm
\vskip3mm
Let us comment upon this result:

{$\star$} {For certain $\e$'s,
there are many successful mass matrices.} 

{$\star$} The value $\e=1$ is not  
particularly good (especially for {triplet}
case). Decreasing $\e,$ the cut on $\theta_{13}$
(CHOOZ) becomes ineffective, and an LMA peak arises.

{$\star$} In the {triplet} 
case, the success takes place for well separated 
set of  values of $\e$; LMA is rather prominent.

{$\star$} In the {seesaw} case, solutions like LOW are often found.
This solution needs a big hierarchy, namely
a little $({\Delta m^2_{sol}}/{\Delta m^2_{atm}})^{1/2},$
and piling-ups of ${\cal O}(1)$ coefficients help 
obtaining that.

{$\star$} {Why there is a correlation  between 
$\e$ and the solar $\nu$ solutions?} 
Let us diagonalize approximatively
the dominant block. The $\nu$ 
mass matrix becomes:
$$
{\bf M_\nu}\propto
\left(
\begin{array}{ccc}
\e^2 & \e & 0 \\
\e & \delta & 0 \\
0 & 0 & 1
\end{array}
\right)
$$
\noindent $\delta$ depends on the 
the dominant block:
it can be little especially for {seesaw} mass mechanism,
because ${\cal O}(1)^{-1}\neq {\cal O}(1)$ 
for matrices. Given $\delta,$ 
SMA prefers certain small values of $\e;$ similarly there is
an optimal value of $\e\sim\delta$ where LMA and LOW 
arise.
\vskip3mm
The question becomes then: \underline{{\sf What is the value of} $\e$?}

\subsection*{More Guesswork \& Some Theory}

(Let us make a step back.)
Froggatt and Nielsen suggested 
that gross structure of {quark} mass matrices 
is ``explained'' by a small\footnote{However, the parameter
$v/M$ should be not {\em too} 
small, if one wants to explain $\sin\theta_C$ itself.} 
ratio
{$v/M$}, and 
a {\em flavor and field dependent} set of charges 
{$Q(q_i)$}  and {$Q(q_i^c)$}
such that:
$$
{\cal L}_{mass}\in q_i q_j^c \times {\cal O}(1)_{ij}\times 
\left(\frac{v}{M}\right)^{Q(q_i)+Q(q^c_j)} +h.c.
$$
It seems we are doing just the same for leptons! Let us buy U(1) 
selection rules.\\ 
Since the charges of {left leptons} are  
almost fixed, only few choices for the {right leptons} 
charges reproduce the correct mass hierarchies.
Optimal values for $v/M=(m_\mu/m_\tau)^{1/2}\sim \sin\theta_C$ are:
\begin{table}[th]
\begin{center}
\begin{tabular}{|ccc|ccc|c|}
\hline
  $Q({\rm e})$ &  $Q(\mu)$ &  $Q(\tau)$ &
  $Q({\rm e^c})$ &  $Q({\mu^c})$ &  $Q({\tau^c})$ &
  $\varepsilon$ (degrees) \\
\hline
3 & 0 & 0 & 3 & 2 & 0 &  $.83^\circ$ \\
2 & 0 & 0 & 4 & 2 & 0 &  $3.4^\circ$ \\
1 & 0 & 0 & 5 & 2 & 0 &  $14.^\circ$ \\
\hline
\end{tabular}
\end{center}
\end{table}

\noindent{The value of {$Q(\mu^c)$} is needed for $m_\mu/m_\tau;$
the sum rule of {$Q({\rm e})+Q({\rm e}^c)=3\times Q(\mu^c)$} is 
needed for $m_{\rm e}/m_\tau\approx (m_\mu/m_\tau)^3.$} Thus, in these
assumptions, we arrive at the striking conclusion that:
\vskip.3cm
\centerline{\fbox{{$\e=(v/M)^{Q({\rm e})}\mbox{ \sf comes 
in quantized values !!!}$}}}
\vskip3mm
Note $Q(\nu_\mu)$ and $Q(\nu_\tau)$ are 
the same--degenerate charges--
that formally is licit, but a bit odd in the {\em spirit} 
of the approach (maybe, neutrinos are really a bit odd).

\subsection*{Implications (Predictions)}
If one takes the point of view of  {Froggatt and Nielsen},
there is a big simplification (in that only certain values
of $\e$ are expected to arise), {but}
there is a part of the analysis above that has to be redone.
\vskip.3cm
In fact, the $\nu$ mixing matrix receives a contribution
from the charged lepton mass matrix:
\begin{equation}
({\bf M_E})_{ij}=(v/M)^{Q(l_i)}\ {\cal O}(1)_{ij}\ (v/M)^{Q(l_j^c)}
\end{equation}
(in other words, in these assumptions the charged lepton mass
matrix is not diagonal in the flavor basis from which we start).
This new contribution to the $\nu$ mixing matrix 
is similar in size to the one of 
the neutrino mass matrix itself; this is not irrelevant,
since, for instance, {$\theta_{13}\sim\e$}
and the probability of survival of electron neutrinos 
in vacuum is:
$$
P_{\rm \nu_e\to\nu_e}\propto \theta_{13}^2
$$
We calculate this new mixing {\em only if} the 
ratios $R_\ell=m_\ell/m_\tau$ are sufficiently well reproduced
$\sum_{\ell={\rm e,\mu}} (R_\ell(th.)/R_\ell(exp.)-1)^2<(30\mbox{\%})^2;$
this happens in $\sim 20$ \% of the cases and permits to avoid 
patological situations.
 \vskip4mm
For more details, check the following table:
\begin{center}
\begin{tabular}{|l|ccccc|}
\hline
$.83^\circ$       & $45-\theta_{23}\!\!\!\!$ 
& $\theta_{13}$ & $\theta_{12}$ & $h$ & 
$m_{\rm ee}/10^{-4}$ \\
\hline
{\tt$\!$t,w/$\!$o}$\!$ & $\!\pm\! 12$ 
      & $\!.37 \!\pm\! .19$ 
      & $\!1.0 \!\pm\! 1.4$ 
      & $\!.35 \!\pm\! .26$           
      & $\!1.4 \!\pm\! 3.3$ 
\\
{\tt$\!$t,w}   & $ \!\pm\! 23$ 
      & $\!.70 \!\pm\! .33$ 
      & $\!1.2 \!\pm\! 1.4$ 
      & $\!.35 \!\pm\! .26$           
      & $\!2.9 \!\pm\! 1.7$ 
\\     
{\tt$\!$s,w/$\!$o}$\!$ & $ \!\pm\! 17$ 
      & $\!.52 \!\pm\! .29$ 
      & $\!1.3 \!\pm\! 1.7$ 
      & $\!.12 \!\pm\! .16$           
      & $\!1.4 \!\pm\! 1.3$ 
\\     
{\tt$\!$s,w}   & $ \!\pm\! 21$ 
      & $\!.79 \!\pm\! .41$ 
      & $\!1.5 \!\pm\! 1.7$ 
      & $\!.12 \!\pm\! .16$           
      & $\!2.9 \!\pm\! 2.4$ 
\\
\hline
\hline
$3.4^\circ$       &  &  &  &  & 
$m_{\rm ee}/10^{-3}$ \\
\hline
{\tt$\!$t,w/$\!$o}$\!$ & $\!\pm\! 12$ 
      & $\!1.5 \!\pm\! 0.8$ 
      & $\!3.8 \!\pm\! 3.8$ 
      & $\!.35 \!\pm\! .26$           
      & $\!2.4 \!\pm\! 0.6$ 
\\
{\tt$\!$t,w}   & $\!\pm\! 23$ 
      & $\!2.9 \!\pm\! 1.4$ 
      & $\!4.6 \!\pm\! 3.8$ 
      & $\!.35 \!\pm\! .26$           
      & $\!4.9 \!\pm\! 2.9$ 
\\     
{\tt$\!$s,w/$\!$o}$\!$ & $\!\pm\! 17$ 
      & $\!2.1 \!\pm\! 1.2$ 
      & $\!5.0 \!\pm\! 5.0$ 
      & $\!.12 \!\pm\! .16$           
      & $\!2.3 \!\pm\! 2.1$ 
\\     
{\tt$\!$s,w}   & $\!\pm\! 21$ 
      & $\!3.3 \!\pm\! 1.7$ 
      & $\!5.7 \!\pm\! 5.1$ 
      & $\!.12 \!\pm\! .16$           
      & $\!4.9 \!\pm\! 4.0$ 
\\
\hline
\hline
$14.^\circ$    
& & & & & 
$m_{\rm ee}/10^{-2}$  \\
\hline
{\tt$\!$t,w/$\!$o}$\!$ & $\!\pm\! 12$ 
      & $\!6.2 \!\pm\! 3.2$ 
      & $\!12.5 \!\pm\! 8.4$ 
      & $\!.36 \!\pm\! .26$           
      & $\!4.0 \!\pm\! 0.9$ 
\\
{\tt$\!$t,w}   & $ \!\pm\! 23$ 
      & $\!11.8 \!\pm\! 5.6$ 
      & $\!16.3 \!\pm\! 9.3$ 
      & $\!.36 \!\pm\! .26$           
      & $\!7.9 \!\pm\! 4.6$ 
\\     
{\tt$\!$s,w/$\!$o}$\!$ & $ \!\pm\! 17$ 
      & $\!8.7 \!\pm\! 4.6$ 
      & $\!17.1 \!\pm\! 12.3$ 
      & $\!.13 \!\pm\! .17$           
      & $\!3.7 \!\pm\! 3.1$ 
\\     
{\tt$\!$s,w}   & $ \!\pm\! 21$ 
      & $\!13.1 \!\pm\! 6.6$ 
      & $\!20.0 \!\pm\! 12.6$ 
      & $\!.13 \!\pm\! .17$           
      & $\!7.6 \!\pm\! 5.9$ 
\\
\hline
\end{tabular}
\end{center}
\vskip.3cm
Here, we show the calculated neutrino properties 
assuming {triplet} or 
{seesaw} ({\tt t} and {\tt s} resp.) mass mechanism, 
and with or without the account of 
the lepton mixing matrix $U_E$ ({\tt w} and {\tt w/$\!$o} resp.).
Note that:

\noindent{$\bullet$} The 3 parts of the table 
correspond to the models defined 
in  previous table (in the left-upper corners, the
values of $\varepsilon$ in degrees are recalled).

\noindent{$\bullet$} Here, 
{$h=\Delta m^2_{sol}/\Delta m^2_{atm}$} and
{$m_{\rm ee}=|({\bf M_\nu})_{\rm ee}|/(\Delta m^2_{atm})^{1/2}$}; 
the angles
{$\theta_{ij}$} are those of neutrino mixing matrix
in the most common (PDG) parameterization. All angles in the table 
are in degrees.

\section{Summary and Discussion}

\noindent{$\star$} We studied  neutrino 
mass matrices with a dominant block and a free parameter,
{$\e$}. This ansatz is motivated by a variety of considerations
(in particular, the large value of {$\theta_{23}$}).

\noindent{$\star$}
Using {random} number generators, we scanned the various 
possibilities and emphasized the most likely outcomes.

\noindent{$\star$} The {\bf triplet} mass mechanism
wants little hierarchy and thence disfavors LOW and 
(less strongly) SMA solutions.
It is more predictive than the {\bf seesaw} mechanism, 
and it likes LMA.

\noindent{$\star$} There is an interesting class 
of mass matrices with
{$\e\sim\sin\theta_C$} (see especially last table). 
They have large {$\theta_{13}$} and give some chance of
success for next generation {$0\nu 2\beta$} experiments, due
to the scaling  {$({\bf M_\nu})_{\rm ee}\propto \e^2.$}

\noindent{$\star$} 
Rotations operating on charged leptons 
(due to U(1) selection rules)
increase {\em (1)} the 
{spread of $\theta_{23}$} 
around $45^\circ$ (unfortunately) 
{\em (2)} the expected {$\theta_{13}$} 
{\em (3)} and 
{$|({\bf M_\nu})_{\rm ee}|= m_{\rm ee}\times (40-70)$ meV.}
\vskip3mm
\noindent In conclusion, 
let us stress that what we presented 
is an appealing  \underline{framework} for massive neutrinos,
more than a compelling theory,
that however--theoretically modest as it is--is
able to give hints for future experiments.
Probably, one should not take these considerations 
{\em too}  seriously; we have a rather limited experimental
information and this makes all too easy to find a successful 
model at present. However, present data certainly point to 
important features of massive neutrinos; 
simple and motivated theoretical proposals 
may help us to delimit  the field of what is 
known, and may perhaps suggest useful new views.
\newpage
\noindent{\large\bf Note to Bibliography}
\vskip3mm
\noindent The present study describes theoretical speculations,
but Section 1 is mostly based on experimental facts: 
\cite{atm,sol,sa,sn,lsnd} 
and little theoretical ingredients \cite{theo}.
The simple minded argument of Section 2 are taken from 
\cite{mg9}; it subtends also \cite{sy98}.
Together with the seminal paper \cite{fn},
the works in \cite{sy98} form the conceptual 
basis of Section 3, where several results of 
\cite{vis01} (the main reference for details
and further information) have been reproduced.
Other relevant works are \cite{hal00}
(the case $\e=1$ and the use of random number generators),
\cite{sy00} (the case $\e=m_\mu/m_\tau$), and \cite{doms}.
A similar but {\em different} class of mass matrices 
\cite{barr} has been denoted  as ``lopsided''; the reason 
is that in these models the neutrino mixing comes mostly 
from ${\bf M_E},$ that is not the case of the models 
considered here.

\end{document}